# Advanced Data Analysis of Spontaneous Biophoton Emission: A Multi-Method Approach

M. Benfatto[*,1], L. De Paolis[*,1], L. Tonello[2,3] and P. Grigolini[3]

[1]Laboratori Nazionali di Frascati, Istituto Nazionale di Fisica Nucleare, Via E. Fermi 40, 00044, Frascati, Italy.
[2]Gioya Higher Education Institution, E305, The Hub Workspace, Triq San Andrija, SGN1612, San Gwann, Malta.
[3]Center for Nonlinear Science, University of North Texas, P.O. Box 311427, Denton, TX 76203-1427, USA.

[*]Corresponding authors: maurizio.benfatto@lnf.infn.it, luca.DePaolis@lnf.infn.it

**Abstract**

Ultra-weak photon emission (UPE) from living systems is widely hypothesized to reflect underlying self-organization and long-range coordination in biological dynamics. However, distinguishing biologically driven correlations from trivial stochastic or instrumental effects requires a robust, multi-method framework. In this work, we establish and benchmark a comprehensive analysis pipeline for photon-count time series, combining Distribution Entropy Analysis, Rényi entropy, Detrended Fluctuation Analysis, its generalization Multifractal Detrended Fluctuation Analysis, and tail-statistics characterization. Surrogate signals constructed from Poisson processes, fractional-Gaussian noise, and renewal processes with power-law waiting times are used to validate sensitivity to memory, intermittency, and multifractality. Across all methods, a coherent hierarchy of dynamical regimes is recovered, demonstrating internal methodological consistency. Application to experimental dark-count data and attenuated coherent-laser emission confirm Poisson-like behavior, establishing an essential statistical baseline for UPE studies. The combined results show that this multi-resolution approach reliably separates trivial photon-counting statistics from structured long-range organization, providing a validated methodological foundation for future biological UPE measurements and their interpretation in the context of non-equilibrium statistical physics, information dynamics, and prospective markers of biological coherence.

**Introduction**

All living systems emit ultra-weak electromagnetic radiation in the visible spectral range, corresponding to approximately 100 photons per second per square centimeter of surface area. This spontaneous emission, universally referred to as biophotons [1,2], has been detected in virtually all aerobic organisms and ceases immediately upon death. Such behavior excludes the possibility that the signal originates from radioactive decay or cosmic ray interactions. Biophotonic emission is characterized by extremely low intensity and by the absence of distinct spectral lines; rather, it displays a nearly flat spectral distribution between 200 and 800 nm, with a subtle maximum in the orange region. The contribution of blackbody radiation at room temperature, estimated using Planck's distribution, is negligible in this spectral range [3]. Notably, environmental or biochemical stress—such as exposure to chemical agents or temperature changes—typically induces a transient amplification of the emission, often by several orders of magnitude, followed by a slow relaxation toward baseline levels according to a non-exponential power-law decay [1,2].

The idea that living organisms produce electromagnetic emissions dates back to the 1920s, when the Russian biologist A. Gurwitsch [4] hypothesized the existence of a weak electromagnetic field responsible for coordinating cell growth and mitotic activity in neighboring tissues. He termed this phenomenon mitogenetic radiation. Although subsequent experimental evidence, such as that provided by Gabor and Reiter [5], supported Gurwitsch's findings, the concept was largely neglected by the scientific community for decades. Roughly thirty years later, Colli and Facchini [6,7] performed the first quantitative measurements of biological photon emission using single-photon-counting photodetectors. Their pioneering results, too, went unnoticed until the 1970s, when F.A. Popp and collaborators [2] initiated a comprehensive experimental and theoretical program aimed at elucidating the physical origin and biological significance of biophoton emission.

While the experimental existence of biophotonic emission is now firmly established, its generation mechanisms and functional roles in biological systems remain open questions. Current models can be broadly classified into two categories [1,2]. The first attributes the emission to stochastic radiative decay processes of molecules previously excited during metabolism; the second postulates the presence of coherent electromagnetic fields within cells that may govern or modulate the observed emission. Both frameworks predict enhanced photon emission under external or internal perturbations, consistent with experimental observations. These explanations are not mutually exclusive, and biophotonic emission may indeed arise from the coexistence of incoherent and coherent processes. Increasing experimental evidence also suggests that biophotons carry biologically relevant information [8–11]: for instance, the emitted light can stimulate cell division in neighboring systems by up to 30%, a phenomenon known as the mitogenetic effect [11–13]. More recently, biophoton emission has attracted attention as a promising non-invasive probe for biological research, with potential applications ranging from toxicology [14] and physiological monitoring [15] to the diagnosis and treatment of diseases, notably cancer [16]. In this context, the temporal evolution of biophoton emission provides a unique window into the dynamical behavior of living matter, especially when analyzed through advanced quantitative methods.

The data consist of discrete photon counts acquired at fixed temporal intervals and stored as a function of the elapsed time since the onset of the experiment. This collection of measurements con-stitutes a time series that encapsulates the temporal dynamics of biophotonic activity. Typically extending over several days of continuous acquisition, the resulting time series exhibit distinct morphological patterns depending on the biological system under investigation. In some cases—such as cell cultures—the emission displays an approximately stationary behaviour with a nearly constant mean photon rate. In contrast, germinating seeds produce highly structured temporal profiles characterized by well-defined modulations, whereas bacterial systems often yield intermittent, burst-like sequences of photon emission events, forming clustered or "bunched" temporal structures. These diverse signatures reflect the intrinsic dynamical organization and metabolic state of the respective living systems. [1,6,7,17].

To extract the underlying statistical structure of spontaneous biophoton emission, our group has primarily employed the Diffusion Entropy Analysis (DEA) method [17], a powerful quantitative approach for assessing complexity in time series data [18,19]. DEA quantifies the scaling index ($\eta$) associated with the diffusion process reconstructed from the experimental signal, following the Kolmogorov framework of complexity [20]. The scaling behavior, evaluated through the Shannon entropy of the corresponding diffusive trajectory, provides a direct measure of deviation from ordinary randomness. Values of $\eta$ differing from the Brownian benchmark ($\eta = 0.5$) signify the presence of long-range temporal correlations and anomalous dynamical organization within the underlying biological system. Our previous studies [17] revealed a persistent deviation from $\eta = 0.5$ throughout germination, reflecting a continuous evolution in the complexity of the system. The initial germination, during the formation of leaves and roots, is dominated by statistical properties typical of the presence of crucial events, while subsequently there is a transition to a phase dominated by properties typical of Fractional Brownian Motion (FBM), probably due to stress on the plant which is measured in a completely light-free environment and therefore cannot begin chlorophyll synthesis for carbohydrate production [21].

These findings indicate that biophoton emission can serve as a quantitative probe of the thermodynamic and informational activity of living systems [2]. Temporal variations in emission intensity correspond to shifts in scaling parameters, offering a dynamic signature of underlying biological transitions and adaptive processes. Within this framework, germination may represent a phase transition in complexity, driven by cellular differentiation and intercellular communication through entropy exchange and information transfer [11].

In recent years, beyond our own investigations, several important contributions have advanced the statistical analysis of ultra-weak photon emission (UPE) from living systems. For example, Michal Cifra [22] and collaborators published a critical review of the photocount-statistics approach in biophoton research. They examine deviations from Poisson behaviour in the photon-count distributions, scrutinise claims of coherence and squeezed states, and emphasise the need for rigorous statistical characterisation of spontaneous biophoton signals.

Similarly, Fritz-Albert Popp and his team laid foundational work in the field of biophotons, emphasising the possibility of coherence in biological photon emission and proposing that statistical properties of the emission (e.g., bursting behaviour, deviations from thermal emission) might carry biologically relevant information [2]. While some of Popp's interpretations remain controversial, his emphasis on statistical features beyond mere intensity (such as time-structure and photon correlations) has inspired much of the subsequent methodological work in this domain.

More recently, two newer studies enrich the methodological toolkit. In 2022, Zsolt Pónya and K. Somfalvi-Tóth [23] analysed UPE kinetics in *Helianthus annuus* plants exposed to biotic and abiotic stress, showing that the decay kinetics of photon emission (as derived from count versus time data) vary significantly with stress types, and fitting these using exponential or power-law models with characteristic time constants. While not strictly a complexity-scaling study, this work illustrates how temporal modelling of photon counts can link statistical parameters to physiological state. In 2024, a broader review by Ganna Nevoit [24] and colleagues explored the role of biophotons in electromagnetic signalling, and included sections on statistical analyses of count distributions, entropy measures, and complexity indices in living systems. These newer approaches underscore the shift toward integrating multiple statistical tools (count distributions, temporal kinetics, complexity scaling) in biophoton research.

The present work aims to provide a comprehensive and comparative analysis of the principal statistical approaches applicable to spontaneous biophoton emission. Our objective is to integrate the different methodologies developed in previous decades, ranging from classical photon-count distributions to modern measures of complexity, into a unified interpretative framework capable of capturing the multifaceted dynamics of living systems.

In particular, we systematically compare and interrelate the information derived from photon-count distribution analyses, Renyi entropies, complexity-scaling techniques, like Diffusion Entropy Analysis (DEA), Detrended Fluctuation Analysis (DFA) and its multifractal generalization (MFDFA). By confronting these complementary perspectives, we aim to identify the distinct statistical signatures that each method reveals, clarify their mutual relationships, and determine how they collectively describe the temporal organization and biological meaning of ultra-weak photon emission. This integrative approach is intended to advance the understanding of how the statistical architecture of biophotonic signals reflects the underlying thermodynamic, metabolic, and informational activity of living matter.

The structure of this work is as follows. In the next section, we present a detailed description of the analytical methods employed, emphasizing their respective conceptual foundations, operational differences, and potential complementarities. Subsequently, we apply these approaches to a set of carefully designed surrogate test series in order to validate their performance and sensitivity in controlled conditions. The analysis then extends to experimental photocounting data, including a characterization of dark noise and of photon-count distributions derived from attenuated laser emission used as reference signals. Finally, we conclude with a discussion of the main findings, their implications for the study of spontaneous photon emission, and prospective directions for future research.

## 2. Experimental data and statistical analysis

Spontaneous biophoton emission is typically measured using a photomultiplier tube (PMT) operating in single photon counting mode. In this configuration, each detected photon event is registered individually, and the total number of counts recorded within a fixed time window constitutes the basic experimental observable. The sequence of photon counts thus obtained forms a time series, whose duration can range from a few hours to several days, depending on the biological system under study. The morphological features of these series differ markedly across systems: for instance, germinating seeds often exhibit highly structured temporal profiles characterized by distinct emission patterns, whereas cell cultures generally display flatter, quasi-stationary trends with comparatively low variability. The detailed behaviour and representative examples of such experimental time series have been thoroughly described in previous studies [17, 25].

In the following section, we present several analytical methods suitable for the quantitative examination of these experimental data, emphasizing their methodological foundations, distinctive features, and potential for revealing the underlying statistical and dynamical properties of spontaneous biophoton emission.

## 2.1. The Probability Distribution Function Approach

Within a semiclassical description of optical detection, a phototube converts the cycle-averaged classical intensity $\bar{I}(t)$ into discrete photocounts. The number of counts m detected in an integration time T is proportional to the incident intensity [26]. From the experimental time series one can construct the photocount distribution $P_m(T)$, i.e., the relative frequency and, upon normalization, the probability, the so-called probability distribution function (PDF), of observing $m$ counts within the acquisition window T. The moments of $P_m(T)$ (mean, variance, higher-order moments) provide diagnostic information on the statistical character of the emission.

Closed-form expressions for $P_m(T)$ are available only for specific physical regimes. The simplest case is a stable classical field with constant cycle-averaged intensity, for which the photocount distribution is Poissonian [22,25,26]:

$$P_m(T) = \frac{\langle m \rangle^m}{m!} e^{(-\langle m \rangle)} \tag{1}$$

Here, $\langle m \rangle$ denotes the mean count (proportional to the mean intensity via a detector-dependent constant), and the variance equals the mean ($\sigma^2 = \langle m \rangle$). Deviations from Poisson statistics are commonly summarized by the Fano factor F defined through $F = \sigma^2/\langle m \rangle$. For a thermal (chaotic) field with $M$ modes, the photocount distribution assumes the negative-binomial form [22,25]:

$$P_m(T, M) = \frac{(m+M-1)!}{m!(M-1)!} \left(1 + \frac{M}{\langle m \rangle}\right)^{-m} \left(1 + \frac{\langle m \rangle}{M}\right)^{-M} \tag{2}$$

In thermal states the mean and variance obey:

$$\sigma^2 = \langle m \rangle + \frac{\langle m \rangle^2}{M} \tag{3}$$

As $M$ becomes large, the variance approaches the mean counts and the thermal distribution tends to the Poisson limit, making discrimination between coherent and thermal emission difficult when many modes are present. In practice, biophoton emission often exhibits super-Poissonian statistics (variance exceeding the mean), consistent with thermal or an integration time T much longer than the characteristic time of the intensity fluctuations of the light beam. This places stringent requirements on experimental approaches aimed at assessing coherence length and time via interference or higher-order correlation measurements.

It is interesting to note that the negative binomial form is obtained from the Poisson distribution by imposing a fluctuation of the average of the gamma type:

$$f(\lambda) = \frac{1}{\Gamma(r)\theta^r} \lambda^{r-1} e^{-\lambda/\theta} \tag{4}$$

and therefore, the photocounts distribution $P_m(T)$ takes the form

$$P_m(T, M) = \int_0^\infty P_m(T) \, f(\lambda) \, d\lambda \tag{5}$$

with $<m> = \lambda = M \cdot \theta$ and $r = M$.

To summarize the pure Poisson distribution is a distribution with a fixed mean count $<m>$, typical of a classical coherent emission, while the negative binomial describes a situation where the mean count fluctuates from interval to interval according to the gamma distribution written above.

An additional aspect of the statistical characterization of biophoton emission concerns the analysis of the tails of the photocount distributions. While the probability distribution functions (PDFs) discussed in the previous section provide global information on the overall variability of the photon counts, a more refined description can be obtained by examining the behavior of their extreme events, i.e., the probability of large fluctuations in the emission intensity. Such analysis offers valuable insight into the stochastic and correlation structure of the underlying processes. In the case of biophoton emission, the observed distributions often exhibit deviations from the ideal Poisson law, displaying heavier tails.

A practical approach to quantify the statistical weight of extreme events is through the definition of tail indices computed [27] for selected quantile thresholds q. These parameters quantify the relative contribution of rare, high-count events to the overall statistical behavior of the system. When applied to photon-count time series, tail analysis provides a complementary measure of variability and complexity, particularly sensitive to nonstationary or burst-like behavior often present in biological emission data. For each experimental PDF $P_m(T)$ we construct the quantity $F_{exp}(k) = \sum_{m \leq k} P_m(T)$ and for a given quantile q (for the tail analysis is convenient to use q ≥ 0.9) we define a threshold $thr_q = min\{k: F_{exp}(k) \geq q\}$, with this we calculate the corresponding $m$ value and later the following right tail indicators:

$$M_{exp}(q) = \sum_{m \geq thr_q} P_m(T) \qquad (6)$$

$$ES_{exp}(q) = \frac{\sum_{m \geq thr_q} m \cdot P_m(T)}{M_{exp}(q)} \qquad (7)$$

The quantity $M_{exp}(q)$ represents the probability of having counts $\geq thr_q$ found with the chosen quantile, while the quantity $ES_{exp}(q)$ represents the conditional average value of the counts that exceed the threshold, i.e. how large the count is in the tail defined by the quantile. From a broader methodological perspective, tail statistics provide a natural link to the nonlinear and complexity-based analyses introduced in the following sections. The presence of heavy tails in the count distributions reveals the dominance of intermittent, non-Gaussian fluctuations—an aspect further quantified by the scaling exponents obtained from DEA and by the multifractal spectra derived from Detrended Fluctuation Analysis, the MFDFA approach. Likewise, rare and intense emission events affect the entropy-based measures, such as the Rényi entropy, by altering the contribution of low-probability fluctuations to the global information balance. In this sense, the analysis of distribution tails can be seen as the starting point for subsequent scaling and entropy approaches capable of investigating complex and self-organizing dynamics.

*2.2. The Diffusion Entropy Approach*

There is now broad evidence that biological systems cannot be adequately described within the ordinary framework of equilibrium statistical mechanics. Living matter inherently violates the assumptions on which Boltzmann's view relies—absence of memory, short-range interactions, and lack of cooperative behavior. Consequently, quantitative tools capable of revealing deviations from equilibrium statistics are essential to characterize the emergent organization of biological processes. Any measurable departure from the canonical equilibrium form can be regarded as a signature of some degree of complexity.

Complexity in this context reflects a delicate balance between order and randomness. When either dominates, the system simplifies; when they coexist in dynamic equilibrium, rich and adaptive behaviors emerge. In germinating seeds, for instance, self-organization is triggered by hydration, transforming a seemingly inert structure into a network of interacting units that collectively express non-linearity, self-similarity, and cooperative dynamics—hallmarks of complex systems.

Within the theoretical landscape of complexity, the concepts of Kolmogorov complexity and Kolmogorov–Sinai entropy provide rigorous measures of the unpredictability and information content of dynamical systems [20,28]. Their evaluation has motivated two major methodological approaches: (i) the compression approach, which seeks to infer Lyapunov exponents or predictability directly from data reduction schemes, and (ii) the diffusion approach, in which the original time series is mapped onto a diffusion-like process

The Diffusion Entropy Analysis belongs to the latter class [28,29,30]. It transforms the information content of a time series into a scaling parameter, the entropy scaling index η, which quantifies the deviation from ordinary Brownian randomness (η = 0.5). Any systematic departure from this value indicates the presence of long-range correlations, memory effects, or cooperative behavior, thus providing a quantitative fingerprint of complexity in living systems. In our context, the original signal is the photon-count series ξ(t), obtained by binning the time axis into intervals of size s, and assigning to each bin the number of detected photons.

Let the original signal be the photon-count time series ξ(t), obtained by dividing the time axis into bins of width s and assigning to each bin the number of photons detected. From time series of length $M$, one constructs the diffusion trajectory
:
$$x(t) = \int_0^t \xi(t')dt' + x(0) \tag{8}$$

For simplicity the time series is considered as a continuous-time series. It is convenient to consider the $x^2(t)$ time series directly related to the correlation function of the original time series [28,29,30]. The scaling properties are determined through the long-time limit behaviour of the correlation function $\langle \xi(t_1)\xi(t_2) \rangle$ and the average can be made over a large number of realizations of $x^2(t)$ using the moving window method [28,29]. Following the standard approach of assuming that the correlation functions are stationary, it is possible to define a normalized correlation function totally independent of the absolute values of $t_1$ and $t_2$:

$$\Phi_\xi(\tau) = \frac{\langle \xi(t_1)\xi(t_2) \rangle}{\langle \xi^2 \rangle} \tag{9}$$

where $\tau = |t_1 - t_2|$ and it is related to the $x^2(t)$ time series by the equation:

$$\langle x^2(t) \rangle = 2 \langle \xi^2 \rangle \int_0^t dt' \int_0^{t'} dt'' \, \Phi_\xi(t'') \tag{10}$$

We can now relate the complexity of ξ(t) to the anomalous scaling of the diffusion trajectory x(t). Using the Fractional Brownian Motion and Hurst notation [30] we indicate the scaling factor with the symbol H. Assuming $x \propto t^H$, differentiation of the second-moment relation (Eq.10) gives the long-time:

$$\Phi_\xi(t) \propto 2H(2H-1) \, t^{2H-2} \tag{11}$$

which has, in the long-time limit the structure $\Phi_\xi(t) \propto \pm \frac{1}{t^\delta}$ with $\delta = 2 - 2H$. Any $H \neq 0.5$ indicates anomalous correlations even in the stationary cases.

Normally we have at our disposal only one time series. To make the statistical analysis we convert the diffusional trajectory $x(t)$ into many realizations so as to make it possible to do an ensemble average. These realizations are performed through a window of size $l$ that we move along the trajectory $x(t)$. Assuming a window of length $l$ ranging from $t$ to $t + l$, the value $x(t)$ can be though as the initial position of the random walker that jumps in a time $l$ from the origin to a value $x = x(t + l) - x(t)$. With a time series of length $M$, realizations of the diffusion process at scale l are obtained via the moving-window definition:

$$x(l, t) = \int_t^{t+l} dt' \, \xi(t') \tag{12}$$

Note that the largest value of $l$ is $M - t$, and for any given window size $l$, we can generate $M - l + 1$ realizations changing the initial position of the random walker, i.e. the t value. From these realizations we estimate the probability density $p(x, l)$ and the related Shannon entropy of our diffusion process [18]. The scaling ansatz

$$p(x.l) = \frac{1}{t^\eta} F\left(\frac{x}{t^\eta}\right) \tag{13}$$

implies for the Shannon entropy

$$S(l) = -\int_{-\infty}^{+\infty} dx \, p(x, l) \ln[p(x, l)] \tag{14}$$

the equation

$$S(l) = A + \eta \cdot \ln(l) \tag{15}$$

This equation means that the entropy $S(l)$ increases linearly with $\ln(l)$ and the slope of the resulting straight line is the scaling factor $\eta$ that must be found numerically from the experimental data.

The numerical results are expressed in a linear-log scale that transform the fitting curve with the form $K + \eta \ln(l)$ into a straight line. Of course, if the FBM condition applies, $\eta = H$.

Although the conversion of a time series $\xi(t)$ into a diffusion trajectory naturally allows one to relate its complexity to a Hurst-like coefficient $H \neq 0.5$, there exists another class of anomalous behaviors that cannot be captured by stationary correlation functions alone [19,31]. A hallmark of many complex systems is the occurrence of *renewal events* [31,32] discrete occurrences that reset the system's memory. The waiting times $\{\tau_i\}$ between successive renewal events are independent and identically distributed random variables, typically following a power–law probability density

$$\psi(\tau) \propto \tau^{-\mu} \qquad 1 < \mu < \infty$$

Such events induce a rejuvenation of the system: each occurrence marks a complete renewal of its internal state. When the exponent $\mu$ lies within $1 < \mu < 3$, the corresponding crucial events produce long–tailed waiting-time distributions that strongly deviate from equilibrium expectations, giving rise to intermittent, burst-like dynamics and anomalous diffusion. A paradigmatic example of a crucial event is the sudden collective reorientation in a flock of birds: when such a maneuver occurs, the global velocity of the swarm transiently vanishes and a new, uncorrelated direction of motion emerges [33]. Analogous mechanisms have been identified in a variety of biological and physiological systems, where crucial events play an essential role in maintaining adaptability and promoting self-organization [34].

In the context of biophoton emission, Diffusion Entropy Analysis can quantify the scaling of complexity but cannot, in its original formulation, discriminate whether an observed anomalous scaling originates from long-range stationary correlations or from renewal–driven intermittency. To resolve this ambiguity, the DEA algorithm can be modified through the introduction of *stripes* [18,34]. In this approach, rather than directly converting the signal $\xi(t)$ into a diffusion trajectory, the amplitude axis is partitioned into contiguous bins (or stripes) of width $\Delta_s$. The times $\{t_i\}$ at which the signal crosses from one stripe to the next are recorded, defining a new event sequence. From this sequence, a binary time series $z(t)$ is constructed as follows. If time $t$ coincides with one of the times $t_i$, we set $z(t) = 1$ and $z(t) = 0$ otherwise. This series $z(t)$ represents a walker that advances by one unit at each event, thus transforming the renewal process into a diffusion trajectory via the same cumulative prescription used for $\xi(t)$. The corresponding entropy $S(\ell)$ is then computed as a function of the window size $\ell$, and its scaling yields the exponent $\eta$, which replaces $H$ in characterizing renewal-driven complexity. It can be shown [17,34] that $\eta$ and $\mu$ are connected through distinct relations depending on the regime:

$$\eta = \mu - 1 \ (1 < \mu < 2), \qquad \eta = \frac{1}{\mu - 1} \ (2 < \mu < 3), \qquad \eta = 0.5, (\mu > 3).$$

When $\mu > 3$, the process corresponds to stationary fluctuations consistent with ordinary equilibrium statistical mechanics. Conversely, $\mu < 3$ signals non–stationary dynamics transient for $2 < \mu < 3$, and permanently out of equilibrium for $1 < \mu < 2$. For the DEA analysis performed without stripes [17,34], we adopt the relation $\mu = 4 - 2\eta$. It is worth noting that, for the limiting cases $\eta = 0.5$ and $\eta = 1$, this expression yields results consistent with those obtained from the general relations previously discussed.

In practical applications, the reliability of the stripes method depends critically on the appropriate choice of stripe amplitude $\Delta_s$. Empirically, we found that setting

$$\Delta_s \approx 3 \cdot (1 \pm \frac{n}{100}) \cdot \sigma,$$

where $\sigma$ is the standard deviation of the original photon count series and $n$ is a number between 0 and 100 to be between $2 \cdot \sigma$ and $4 \cdot \sigma$, ensures optimal discrimination of renewal events. This rule provides a balance between event density and statistical independence of crossings and has been validated on surrogate data sets with known statistical properties. The recurrence of this *three–sigma principle* is noteworthy: although the DEA operates far beyond the Gaussian domain, choosing $\Delta_s$ near a few standard deviations naturally preserves the relevant dynamical range of fluctuations while suppressing noise–induced micro–crossings [27].

The practical determination of the scaling exponent $\eta$ is done through a linear fit of $S(\ell)$ as a function of the logarithm of the window length $\ell$. One of the main practical challenges in DEA concerns the selection of the fitting interval, since statistical fluctuations can significantly affect entropy values near the boundaries. To obtain a robust characterization of the scaling exponent and to evaluate its local stability, a *sliding–window fitting procedure* was implemented. The method operates over the global range $[\log(\ell_{\min}), \log(\ell_{\max})]$ of the DEA curve and proceeds as follows:

1. A moving window of width $w$, containing $n$ evenly spaced data points, is defined along the logarithmic axis.
2. Within each window, a local linear regression of $S(\ell)$ versus $\log(\ell)$ is performed, yielding a local slope $\eta_i$ and its uncertainty $\Delta_{\eta_i}$.
3. The window is then shifted by a step length $l$, generating a sequence of local slope estimates

$$\{\eta_1, \eta_2, \eta_3, \ldots, \eta_k\}.$$

For each contiguous block of slopes of fixed length (set to five windows in our implementation), the standard deviation $\Delta_\eta$ of the corresponding slopes is computed. A block is considered *stable* if $\Delta_\eta < \varepsilon$, with the empirical threshold $\varepsilon = 0.02$. When a stable plateau is identified, the mean scaling exponent and its uncertainty are calculated as weighted averages:

$$\eta_{\text{plateau}} = \frac{\sum_{i \in \text{plateau}} \eta_i / \Delta_{\eta_i}^2}{\sum_{i \in \text{plateau}} 1/\Delta_{\eta_i}^2}, \qquad \Delta_{\text{plateau}} = (\sum_{i \in \text{plateau}} \frac{1}{\Delta_{\eta_i}^2})^{-1/2}.$$

These values represent the weighted mean and associated standard deviation of the locally stable scaling region. If no plateau satisfies the stability criterion, the algorithm concludes that no statistically reliable region is present within the selected interval. Conversely, when a plateau is detected, the corresponding $\eta_{\text{plateau}}$ is used to perform a constrained global fit of $S(\ell)$, fixing the slope to the plateau value. This dual local–global approach allows for direct comparison between local and global scaling behaviors, revealing potential transitions or instabilities in the entropy dynamics. At the end of the analysis, the final scaling exponent $\eta$ is determined by averaging within the identified plateau region, where the variability of the slope is minimal. This ensures a robust estimation of the scaling regime even under experimental noise.

*2.3. The Rényi Entropy*

In the quantitative characterization of complex biological dynamics, the use of generalized entropic measures provides a powerful extension of the conventional Shannon framework. Among them, the Rényi entropy $H_\alpha$ offers a flexible approach to probe the statistical structure of a system across multiple scales of probability weighting [35,36]. For a discrete probability distribution $\{p_i\}$, the Rényi entropy is defined as

$$H_\alpha = \frac{1}{1-\alpha} \ln \left( \sum_i p_i^\alpha \right),$$

where $\alpha \in \mathbb{R}$ is the order of the entropy. In the limit $\alpha \to 1$, $H_\alpha$ reduces to the Shannon entropy. The parameter $\alpha$ acts as a statistical "lens," enhancing or suppressing contributions from different regions of the probability distribution. For $\alpha > 1$, entropy becomes increasingly sensitive to high-probability events, while for $\alpha < 1$ (and especially for negative values) it emphasizes the contribution of rare or extreme fluctuations. This property makes the Rényi entropy particularly suitable for biological time series, where the coexistence of long-range organization and rare bursts of activity reflects the non-equilibrium and self-organizing nature of living matter. From a conceptual standpoint, the Rényi entropy provides a bridge between statistical mechanics and information theory, allowing one to quantify deviations from canonical equilibrium behavior. In biological systems, where the assumptions of short-range interactions and ergodicity are often violated, the dependence of $H_\alpha$ on the entropic order becomes an indicator of the system's hierarchical structure and the diversity of dynamical states explored over time.

Two complementary computational approaches were adopted to estimate $H_\alpha$ from the biophoton emission time series.

(i) *Direct estimation*:

The direct method evaluates the probability density function of the experimental signal by means of a *kernel density estimation* (KDE) technique with reflective boundaries on the normalized domain [0,1]. This approach reconstructs the continuous probability landscape of photon counts, thereby allowing the computation of $H_\alpha$ directly from the empirical PDF:

$$H_\alpha^{(\text{dir})} = \frac{1}{1-\alpha} \ln \left( \int_0^1 [p(x)]^\alpha \, dx \right).$$

The normalization $\widetilde{H}_\alpha = H_\alpha / \ln K_{\text{dir}}$, where $K_{\text{dir}}$ is the number of KDE bins, ensures comparability across datasets with different sampling resolutions. This formulation captures the global variability and amplitude distribution of the photon signal, making it sensitive to broadening of the distribution tails and to transitions between stationary and intermittent regimes.

(ii) *Symbolic sequence estimation*:

The symbolic method operates on a transformed representation of the signal, where the original continuous data $x(t)$ are converted into a sequence of discrete symbols [37,38]. This procedure, rooted in *symbolic dynamics* and *n-gram analysis*, aims to capture not only the statistical dispersion of the values but also the sequential order of events, thus incorporating temporal correlations that are invisible to purely amplitude-based approaches. The transformation proceeds in two steps. The first one is a kind of quantization or symbolic encoding which depends on the nature of the original data.

For continuous-valued signals, the transformation is achieved through a quantile-based discretization of the amplitude distribution. The full range of values of $x(t)$ is partitioned into $K_{seq}$ equiprobable intervals (quantile bins), each corresponding to one symbol in the discrete alphabet $\{0, 1, \ldots, K_{seq} - 1\}$. In this representation, each symbol denotes a *set* of numerical values — a quantile-defined range, rather than a single amplitude. This procedure ensures that all symbols have comparable marginal probabilities and that the resulting symbolic sequence $s(t)$ retains the statistical structure and ordering of the original signal while being invariant to its absolute scaling.

For discrete-valued signals, such as integer or quasi-integer time series (e.g., photon counts or digital levels), no quantization step is required. Each distinct value of $x(t)$ naturally defines a unique symbol, so that $s(t) = x(t)$ and the alphabet size $K_{seq}$ equals the number of distinct observed values. In this case, the symbolic representation directly reflects the empirical states of the system, and the subsequent $n$-gram construction captures the recurrence and ordering of these states over time.

The second step is the calculation of the Renyi entropy. From the symbolic sequence, contiguous words of length $L_{seq}$ are extracted to build the $n$-gram probability distribution. The symbolic space explored by the sequence-based analysis is fully defined by the parameters $K_{seq}$ and $L_{seq}$ and the total number of possible symbolic combinations is $K_{seq}^{L_{seq}}$.

The frequency of each unique symbolic sequence defines a discrete probability distribution $\{p_j\}$ over the possible combinations, from which the Rényi entropy is calculated as

$$H_\alpha^{(seq)} = \frac{1}{(1-\alpha)L_{seq}} \ln \left( \sum_j p_j^\alpha \right)$$

and is further normalized with respect to the to the effective alphabet of the series, i.e., the number of distinct observed symbols $K_{seq}$, by dividing by $\ln K_{seq}$. This normalization allows direct comparison across different experimental conditions, segmentations, or alphabet sizes.

Given a series of $N$ samples, the number of observed contiguous symbolic words of length $L_{seq}$ is approximately $N_{obs} = N - L_{seq} + 1 \approx N$. The ratio $N_{obs}/K_{seq}^{L_{seq}}$ quantifies the *average coverage* of the symbolic space, i.e., the expected number of realizations per possible word, and determines the statistical reliability of the entropy estimation. Assuming a uniform distribution, the expected average number of realizations per possible word is therefore

$$\langle n \rangle = \frac{N}{K_{seq}^{L_{seq}}}$$

For instance, with $K_{seq} = 16$, $L_{seq} = 3$, and $N = 10^5$, each 3-symbol word is expected to occur about

24 times on average, ensuring robust statistical coverage of the symbolic space while maintaining sensitivity to short-range temporal correlations.

This symbolic formulation captures the temporal organization of the signal, specifically how patterns of photon emission recur or reorganize over time. When combined with the direct method, it reveals how statistical structure and temporal sequencing contribute differently to the system's overall complexity. From a conceptual standpoint, the symbolic formulation highlights the *temporal grammar* of the signal, the way in which short subsequences of events combine to form higher-order patterns. A reduction in symbolic entropy compared to the direct estimate ($H_\alpha^{(\text{seq})} < H_\alpha^{(\text{dir})}$) indicates that the temporal ordering imposes additional constraints on the system's dynamics, suggesting the presence of memory and self-organization. Conversely, similar values of $H_\alpha$ for both methods suggest that the system behaves more like a random or weakly correlated process.

## 2.4. The DFA and MFDFA approaches

The analysis of spontaneous photon emission time series requires methods capable of detecting long-range correlations under conditions of nonstationary and intermittent dynamics. Ordinary statistical measures, which rely on fixed mean and variance, are inadequate when the signal is organized in bursts and quiescent intervals of variable intensity. In this context, the Detrended Fluctuation Analysis and its multifractal generalization provide a natural framework to quantify scale-dependent organization and to distinguish between random fluctuations and self-organized temporal structures [39,40,41].

In DFA, the original signal $x(k)$ is first converted into its integrated profile

$$Y(i) = \sum_{k=1}^{i} [x(k) - \langle x \rangle],$$

which emphasizes slow components and cumulative correlations. For a scale $s$, the profile is split into $N_s = \lfloor N/s \rfloor$ non-overlapping segments; to reduce edge effects one usually processes both the forward and the reversed series, yielding $2N_s$ segments in total. In each segment $v$ we remove a polynomial trend of order $m$ (typically $m = 1$ or $2$) and compute the local variance

$$F^2(v, s) = \frac{1}{s} \sum_{i=1}^{s} [Y((v-1)s + i) - P_v(i)]^2$$

where $P_v(i)$ is the fitted polynomial. The root–mean–square fluctuation is

$$F(s) = \{\frac{1}{2N_s} \sum_{v=1}^{2N_s} F^2(v,s)\}^{1/2} \sim s^H$$

The Hurst exponent $H$ classifies temporal dependence: $H = 0.5$ (uncorrelated/Poisson-like), $H > 0.5$ (persistent), $H < 0.5$ (anti-persistent).

This approach could be particularly meaningful for biophoton data, where the temporal organization reflects the interplay between stochastic emission and coherent self-organized processes within the living system. Deviations from $H = 0.5$ express the presence of memory and cooperative activity, which are expected in the dynamics of biological growth and adaptation. Moreover, by computing the local slope of $\log F(s)$ versus $\log s$, one can reveal crossovers between scaling regimes, indicating transitions in the internal organization of the system as it evolves.

The multifractal extension [41], MFDFA, provides a more detailed picture by weighting fluctuations according to their amplitude. The $q$-order fluctuation function is

$$F_q(s) = \{\frac{1}{2N_s}\sum_{v=1}^{2N_s}[F^2(v,s)]^{q/2}\}^{1/q}, q \in \mathbb{R}$$

with the geometric-mean definition for $q = 0$:

$$F_0(s) = \exp\{\frac{1}{4N_s}\sum_{v=1}^{2N_s}\ln[F^2(v,s)]\}$$

Positive $q$ amplifies large deviations (bursts), negative $q$ emphasizes small deviations (laminar/quiescent phases). The scaling law

$$F_q(s) \sim s^{h(q)}$$

defines the generalized Hurst exponent $h(q)$. For monofractal $h(q)$ is flat; for multifractals $h(q)$ varies with $q$.

A convenient summary is given by the mass exponent

$$\tau(q) = q\,h(q) - 1,$$

which is the scaling exponent of the partition function $Z_q(s) = \sum_v[F^2(v,s)]^{q/2} \sim s^{\tau(q)}$. Conceptually, $\tau(q)$ tells how the $q$-weighted "mass" of fluctuations redistributes across scales: linear $\tau(q)$ indicates a single scaling law; curvature signals a spectrum of coexisting laws. Through a Legendre transform,

$$\alpha = \frac{d\tau}{dq}, \quad f(\alpha) = q\,\alpha - \tau(q),$$

one obtains the singularity spectrum. Here $\alpha$ is the local Hölder (singularity) exponent while $f(\alpha)$ estimates the fractal dimension of the set of points that share a given $\alpha$. In practice, the peak location $\alpha_{\max}$ reflects the dominant scaling (analogous to a global $H$), and the width $\Delta\alpha = \alpha_{\max} - \alpha_{\min}$ quantifies the degree of multifractality: broader spectra imply stronger heterogeneity of scaling behaviours. Asymmetries of $f(\alpha)$ are informative: a left-skewed spectrum (extended towards small $\alpha$) reveals dominance of intense bursts; a right-skewed spectrum indicates prevalence of quiet, fine-scale fluctuations.

Applied to our emission series, DFA and MFDFA reveal the presence of persistent temporal correlations and multiscale intermittency. The scaling exponent $H$ captures the global tendency toward memory and organization, while the spectrum $f(\alpha)$ highlights the coexistence of multiple dynamical components—ranging from quiet, noise-like intervals to rare and intense bursts. This dual description is crucial for biological systems, where the emission process alternates between phases of quasi-stationary activity and self-organized fluctuations associated with metabolic or developmental transitions. In the present analysis, the same set of scales is used for both DFA and MFDFA, ensuring direct correspondence between the global and the multifractal indicators.

## 2.5. General considerations

The four analysis approaches employed in this study form a coherent methodological framework designed to capture complementary aspects of complexity in biophoton emission. Although they originate from distinct theoretical frameworks, they share a unifying goal: the identification of scale-invariant structures and statistical regularities that depart from equilibrium randomness. Each method isolates a specific facet of the signal—its amplitude distribution, informational structure, or temporal organization—yet all contribute to a common picture in which stochastic variability and self-organized dynamics coexist across multiple time scales.

Tail analysis provides the most direct statistical characterization of photon emission, focusing on the probability distribution of counts and its extreme fluctuations. The indices of tail heaviness quantify the relative weight of rare, high-intensity bursts compared to the background emission. Strongly non-Gaussian tails indicate that the signal is dominated by intermittent activity rather than uncorrelated noise, a hallmark of nonequilibrium behavior in living systems. This amplitude-based approach thus establishes the empirical evidence of intermittency that motivates the application of scaling and entropy-based analyses.

The Diffusion Entropy Analysis translates these amplitude irregularities into an information-theoretic framework. By mapping the photocount series onto a diffusion process, DEA measures how Shannon entropy grows with time window size. The slope of this relationship defines the diffusion exponent $\eta$, which quantifies the rate of information production and reveals whether the system follows normal diffusion ($\eta = 0.5$) or exhibits superdiffusive or sub diffusive dynamics. DEA is particularly suited to detect nonstationary and renewal-driven behavior: it captures memory effects and collective organization even when correlations are not expressible through stationary functions. Unlike fluctuation-based methods such as DFA and MFDFA, DEA does not require detrending or amplitude normalization—which can alter the scaling of fluctuations—but instead derives a distribution-independent measure of information growth and complexity scaling.

The Rényi entropy framework generalizes the concept of information scaling by introducing an order parameter $\alpha$ that modulates the contribution of frequent and rare events. The direct formulation, based on the empirical amplitude distribution, highlights how information is distributed across fluctuation magnitudes; the symbolic formulation, in turn, operates on discretized sequences, making it sensitive to temporal ordering and pattern recurrence. Together, these complementary forms describe how the system allocates information among amplitude and temporal domains.

The Detrended Fluctuation Analysis and its multifractal extension focus explicitly on the temporal correlations within the signal. DFA quantifies how detrended variance scales with time, yielding the Hurst exponent H, which distinguishes between uncorrelated (H = 0.5), persistent (H > 0.5), and antipersistent (H < 0.5) behaviors. MFDFA extends this to a family of generalized exponents h(q), describing how fluctuations of different intensities scale differently, thus revealing the multifractal hierarchy of the signal. In stationary, monofractal systems, $\eta$ from DEA and $H \approx h(2)$ provides equivalent measures of global scaling. However, when intermittency and multifractality dominate—as often occurs in biological photon emission—the two approaches diverge: DEA captures entropy-based nonstationarity and renewal dynamics, while DFA/MFDFA quantify the persistence and heterogeneity of temporal correlations. Their joint interpretation therefore separates temporal memory from statistical nonstationarity, both of which contribute to biological complexity.

Despite their distinct formulations, all methods converge toward a unified description of the system's scaling architecture. The heaviness of the distribution tails signals the presence of strong, rare fluctuations; DEA interprets these as enhanced entropy growth and information production; Rényi entropy quantifies how such events reshape the informational balance between order and randomness; and DFA/MFDFA reveal how these bursts propagate through time, sustaining long-range correlations and multifractal organization. Together, these results show that biophoton emission is not a stationary random process but the manifestation of a self-organized system operating near a critical balance between coherence and variability. The correspondence among the principal scaling parameters, $H$, $h(2)$, and $\eta$ and the similar asymmetries in the Rényi and MFDFA spectra confirm that both informational and dynamical dimensions of the signal obey the same scaling principles. Complexity in this context emerges not from randomness alone but from the structured variability inherent to living matter [42,43], where fluctuations act as a medium of coherence and regulation.

In the following section, we will apply these analytical methods to surrogate time series constructed to reproduce well-defined statistical properties. This will allow us to verify the consistency and sensitivity of each method before extending the analysis to experimental biophoton data.

### 3.0. The application to surrogate series

To validate the statistical methods and benchmark their sensitivity to distinct forms of temporal organization, we generated three classes of surrogate signals: crucial event (CE) sequences, fractional Gaussian Noise (FGN) series, and Poisson noise. It should be noted that FBM is obtained as the time-integral of FGN. This guarantees a consistent representation of long-range correlated Gaussian dynamics. These surrogates provide controlled examples of non-stationary renewal dynamics, correlated Gaussian fluctuations, and purely random emission, respectively—each representing a fundamental dynamical archetype relevant to the phenomenology of biophoton signals.

The crucial event time series [18,44] was generated according to a renewal process characterized by a power-law distribution of waiting times between successive events. The waiting times $\tau$ are produced using the transformation

$$\tau = T\left(\frac{1}{y^{\frac{1}{\mu-1}}} - 1\right)$$

where $y$ is a real number uniformly distributed in the interval $(0,1)$, $T$ is a characteristic time scale, and $\mu$ is the power-law exponent controlling the heaviness of the waiting-time tail. For $\mu > 2$, the average waiting time is finite and given theoretically by

$$\langle \tau \rangle = \frac{T}{\mu - 2}.$$

For $1 < \mu < 3$, the resulting dynamics are non-stationary and characterized by long-term memory arising from the divergence of the second moment of $\psi(\tau)$. The discrete binary series $x(t)$ is obtained by setting $x(t_i) = 1$ at event times $t_i = \sum_{j=1}^{i} \tau_j$ and $x(t) = 0$ elsewhere, thus mimicking a point process with intermittent bursts and laminar intervals. This model captures the renewal nature and self-organized intermittency often observed in biological systems, from neuronal spiking to photon emission [34].

In contrast, fractional Gaussian Noise (FGN) motion represents a Gaussian, self-affine process with stationary increments and long-range correlations governed by the Hurst exponent $0 < H < 1$. FGN was generated using the Davies–Harte [45] embedding method, which ensures exact covariance reproduction in the frequency domain through a circulant-matrix formulation. For $H > 0.5$, the process exhibits persistent correlations (superdiffusion), while for $H < 0.5$ the dynamics become antipersistent (sub diffusion). FGN surrogates provide a reference for correlated but stationary scaling behaviour, enabling direct comparison with the renewal-driven, non-stationary properties of crucial event sequences.

Together, these two surrogate classes establish a complementary framework for testing the performance and interpretive scope of the adopted analytical tools. The CE series introduce heavy-tailed, non-stationary scaling typical of self-organized critical phenomena, whereas the FGN surrogates reproduce correlated yet Gaussian fluctuations with tunable persistence.

Finally, a Poisson surrogate series was generated to represent the null hypothesis of purely random emission. Together, these three surrogate models span the principal regimes of stochastic organization, Poisson (memoryless randomness), FGN (stationary long-range correlation), and CE (non-stationary renewal dynamics)—providing a comprehensive framework for testing the sensitivity and interpretive scope of the analytical methods applied in this study.

The CE series was produced using a renewal process with parameters $\mu = 2.5$ and $T = 1$; the FGN series was generated with a Hurst exponent $H = 0.7$; and the Poisson series was simulated with an average photon count rate of $\langle m \rangle = 10$. The time step used in the generation has been fixed at 1 sec.

Figure 1 shows a graphical representation of the three series over the first hundreds of points (each sequence consists of 100,000 data points in total) to have a better graphical representation. The distinct statistical and temporal characteristics of the CE, FGN, and Poisson processes are evident: the CE series displays intermittent bursts and long waiting times, the FGN series exhibits correlated fluctuations, while the Poisson series represents a purely random baseline.

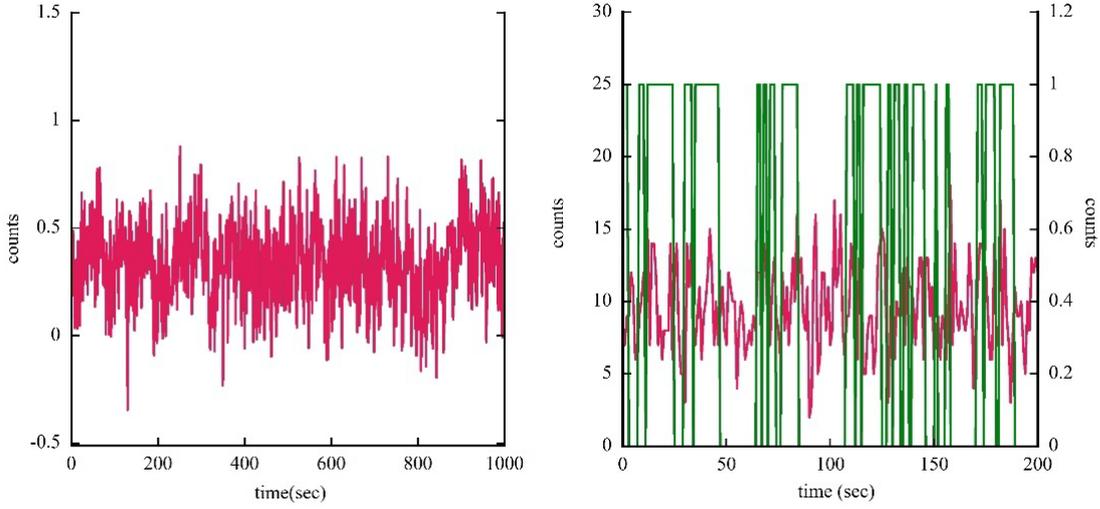

**Figure 1.** *Left:* first 1000 points of the FGN series ($H = 0.7$), composed of real-valued data that can assume both positive and negative values. *Right:* first 200 points of the Poisson (red) and Crucial Event (CE) (green) series.

To better approximate realistic experimental conditions, where biophoton emission typically arises from the superposition of multiple dynamical sources, three additional composite series were constructed by combining the prototype signals with appropriate weighting factors. These hybrid test series reproduce different mixtures of stochastic and organized components and were used for the validation tests presented in this section.

Table 1 summarizes the main characteristics of the reference and composite series, here identified as: P-type (pure Poisson process), C-type (Poisson + CE), F-type (Poisson + FGN component), and CF-type (Poisson + CE + FGN component).

|         | $<m>$ | $\sigma$ | Skewness |
|---------|-------|----------|----------|
| P-type  | 10.0  | 3.17     | 0.31     |
| C-type  | 11.3  | 3.51     | 0.24     |
| F-type  | 11.1  | 3.22     | 0.29     |
| CF-type | 12.5  | 3.56     | 0.23     |

**Table 1**. Main statistical parameters of the four prototype and mixed series. The $\langle m \rangle$ value denotes the mean photon count, $\sigma$ the standard deviation, and *Skewness* the asymmetry of the probability distribution counts.

The values show that all series have comparable variability ($\sigma \approx 3\text{–}3.6$) and weakly positive skewness, indicating slightly asymmetric distributions dominated by small-amplitude events with occasional higher counts. The composite series (CF-type) exhibits a modest increase in both mean and dispersion, reflecting the combined contribution of correlated (FGN-type) and intermittent (CE-type) dynamics to the underlying Poisson background.

If we consider the Poissonian part of the series as noise, we find that the mixed series have a signal-to-noise ratio of the order of $0.5 - 1$, which are typical values in biophotonic signals coming from cell cultures [27].

The first analysis focuses on the probability distribution of photon counts and on the statistical characterization of their tails. Figure 2 illustrates a representative example obtained from the C-type series, together with the position of the selected quantile used for tail evaluation. The other series yield qualitatively similar distributions and are therefore omitted here for brevity.

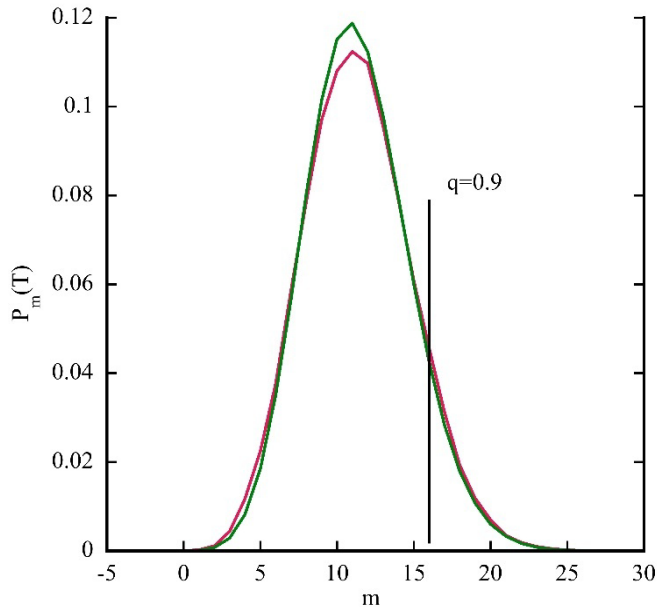

**Figure 2.** Probability distribution $P_m(T)$ for the C-type series (red line) compared with the corresponding Poisson fit (green line). The vertical line indicates the position of the quantile $q = 0.9$ used for tail analysis.

The fit reproduces well the central region of the distribution, while slight deviations appear in the tail, indicating the presence of non-Poissonian fluctuations and a moderate excess of large-count events. The vertical line marks the quantile $q = 0.9$ used for tail analysis. Similar behavior has been found for the other series. All relevant numerical parameters derived from these distributions, including fitted values and tail indicators, are summarized in Table 2 for all examined series.

|  | $M_{exp}(q)$ | $ES_{exp}(q)$ | Fano factor |
| --- | --- | --- | --- |
| P-type | 0.137 | 15.38 | 1.006 |
| C-type | 0.122 | 17.44 | 1.088 |
| F-type | 0.150 | 16.43 | 0.941 |
| CF-type | 0.133 | 18.50 | 1.022 |

**Table 2**. Main parameters derived from the tail analysis of the count distributions. The quantities $M_{\exp}(q)$, $ES_{\exp}(q)$ and the Fano factor are defined in the previous paragraph.

The results reported in Table 2 clearly indicate that while the Fano factor remains approximately constant and close to unity for all series, indicating that the overall fluctuation level is nearly Poissonian, the tail parameters $M_{\exp}(q)$ and $ES_{\exp}(q)$ exhibit clear variations. This suggests that deviations from pure Poisson behavior emerge primarily in the tails rather than in the bulk of the distribution. In fact, fits performed with a thermal (gamma) distribution on the mixed series yield a high number of modes and are practically indistinguishable from a pure Poisson fit, in agreement with previous findings by Cifra et al. [22]. However, the tail analysis highlights subtle but significant departures from the canonical form, offering a more sensitive indicator of the underlying dynamical heterogeneity.

The differences observed in the tail behavior of the count distributions suggest that, beyond simple statistical variability, the underlying emission dynamics may involve distinct temporal organization mechanisms. To investigate whether these amplitude fluctuations correspond to genuine temporal correlations or renewal-like structures, we now turn to the Diffusion Entropy Analysis, which provides a direct measure of scaling and memory effects in the photon-count time series.

Figure 3 illustrates the typical dependence of the Shannon entropy $S(l)$ on the window length $l$, obtained from the Diffusion Entropy Analysis of the C-type series without the use of stripes. The solid line represents the best linear fit derived through the sliding-window procedure described earlier, providing an estimate of the scaling exponent $\eta$ that characterizes the diffusion regime of the signal.

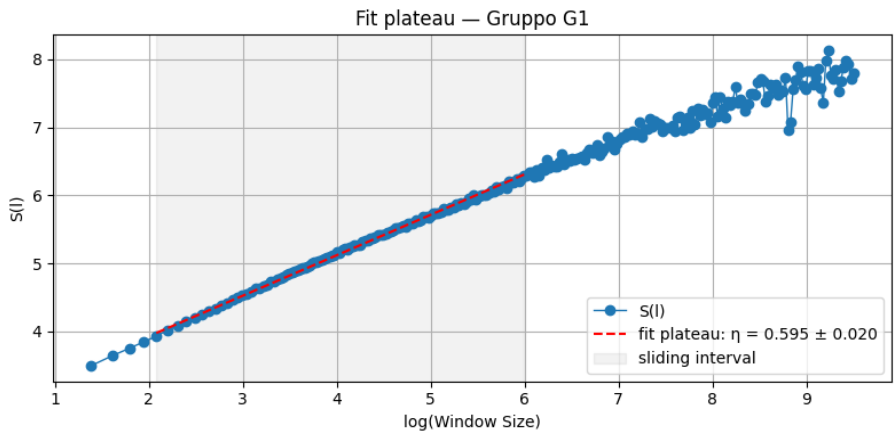

**Figure 3.** Diffusion Entropy Analysis of the C-type series without stripes. The blue points represent the Shannon entropy $S(l)$ as a function of the logarithm of the window size $l$. The red dashed line indicates the best linear fit obtained through the sliding-window procedure in the range $\ln(l) = 2–6$, yielding a scaling exponent $\eta = 0.595 \pm 0.020$.

Similar figures are obtained for all DEA analyses performed here. Table 3 summarizes the results of the Diffusion Entropy Analysis performed on the four surrogate series described above: a purely Poissonian sequence (P-type), a Poisson process modulated by crucial events with power-law waiting times (C-type, µ = 2.5), a Poisson process superimposed with a fractional Gaussian noise (F-type, H = 0.7), and a mixed process combining both crucial events and FGN (CF-type). For each series, the scaling exponent η has been computed both without and with the stripe procedure, using a stripe size $\Delta_s$ equal to 3·σ, as well as slightly larger ($\Delta_s$ = 3.9·σ) and smaller ($\Delta_s$ = 2.7·σ) values to test the robustness of the results. The results confirm that, for a purely Poissonian process, the DEA yields η ≈ 0.5 as expected for a purely stochastic and memoryless dynamics, largely independent of the precise stripe size. In the C-type series, the addition of crucial events produces a higher scaling (η ≈ 0.60 without stripes), reflecting the emergence of long-range correlations and intermittent behavior. The introduction of stripes reduces the scaling to values around 0.55–0.56, consistent with the expected filtering of the deterministic temporal clustering generated by crucial events.

|  | $\eta$ (no-str) | $\eta\ [\Delta_s = 3 \cdot \sigma]$ | $\eta\ [\Delta_s = (3 + 0.9) \cdot \sigma]$ | $\eta\ [\Delta_s = (3 - 0.3) \cdot \sigma]$ |
|---|---|---|---|---|
| P-type | 0.46±0.02 | 0.50±0.02 | 0.49±0.02 | 0.51±0.02 |
| C-type | 0.60±0.02 | 0.56±0.02 | 0.63±0.02 | 0.55±0.02 |
| F-type | 0.54±0.01 | 0.52±0.02 | 0.50±0.02 | 0.50±0.02 |
| CF-type | 0.66±0.01 | 0.59±0.02 | 0.59±0.01 | 0.55±0.02 |

**Table3.** Scaling exponents η obtained from the Diffusion Entropy Analysis (DEA) applied to four surrogate time series: purely Poissonian (P-type), Poisson plus crucial events (C-type, µ = 2.5), Poisson plus fractional Gaussian noise (F-type, H = 0.7), and Poisson plus crucial events plus FGN (CF-type). Results are reported both without stripes and with stripe widths $\Delta_s$ equal to 3·σ, 3.9·σ, and 2.7·σ. The error bars represent standard deviations estimated from linear regression over the selected scaling region.

The F-type process, obtained by superimposing a fractional Gaussian noise with H = 0.7 on a Poisson sequence, shows η ≈ 0.54 without stripes and η ≈ 0.50–0.52 when stripes are applied. This behaviour indicates that the stripe procedure effectively removes most of the fractional component of the dynamics, restoring a nearly diffusive scaling (η ≈ 0.5). Finally, in the CF-type case, where both fractional memory and crucial intermittency are present, the DEA yields η ≈ 0.66 without stripes and η ≈ 0.55–0.59 with stripes, showing an intermediate behaviour between the C-type and F-type signals. This confirms that the stripe-based DEA can selectively suppress the long-memory components of the signal while still preserving signatures of crucial intermittency.

It is worth noting that the analysis of the *pure* surrogate components yields η ≈ 0.77 (no stripes) and η ≈ 0.66 (with stripes) for the crucial event sequence, and η ≈ 0.70 (no stripes) and η ≈ 0.51 (with stripes) for the pure FGN ($H = 0.7$). These values are fully consistent with theoretical expectations: for FGN, the scaling exponent is expected to match the Hurst parameter (η ≈ H), while for crucial events η is theoretically linked to µ through the relation $\eta = 1/(\mu - 1)$, giving η ≈ 0.67 for $\mu = 2.5$. Good agreement between the DEA-derived scaling and theoretical predictions confirms the reliability of the numerical implementation and the physical meaning of the DEA scaling exponent as a marker of the underlying temporal correlations [18]. The deviations observed in the mixed cases are consistent with the competition between fractional persistence and intermittent renewal dynamics, both contributing to the effective scaling measured by the DEA.

Furthermore, it is observed that when the stripe size is reduced too much in the mixed series involving crucial events, part of the underlying stochastic fluctuations becomes misinterpreted by the DEA as spurious crucial events. As a consequence, the scaling exponent η decreases toward the Poissonian value ($\eta \approx 0.5$), indicating that overly narrow stripes cause the analysis to absorb noise contributions as intermittent events, thereby masking the true renewal dynamics. This effect highlights the importance of selecting an appropriate stripe width to correctly separate stochastic variability from genuine crucial dynamics. Empirically, a stripe size in the range between $3 \cdot \sigma$ and $4 \cdot \sigma$ appears to provide a reliable balance, efficiently filtering stochastic noise while preserving the statistical signatures of crucial events.

While the DEA provides a robust assessment of the scaling properties and temporal correlations of diffusion-like processes, it mainly quantifies the global growth of entropy as a function of time. To achieve a deeper understanding of the statistical composition of the signal, we extended the analysis to the Rényi entropy, which generalizes the Shannon measure by introducing an order parameter α. By varying α, different regions of the probability distribution can be selectively emphasized, thus allowing one to distinguish the contribution of rare intermittent events from that of frequent fluctuations. This complementary approach reveals additional layers of complexity beyond the global scaling captured by the DEA.

In the following, the Rényi entropy is evaluated using two complementary approaches. The direct method estimates the probability distribution from the amplitude statistics of the time series, providing a continuous characterization of the signal's variability. In contrast, the symbolic method is based on the discretization of the time series into symbolic sequences, thereby emphasizing the temporal organization of fluctuations rather than their absolute magnitude.

In practice, both entropy formulations were applied to the same time-series under identical conditions. Each photon-emission record $x(t)$ was normalized to the interval $[0,1]$ and, when necessary, divided into contiguous groups of equal length to test the local stability of the results. The Rényi entropy $H_\alpha$ was then computed over a continuous grid of $\alpha$ values, typically spanning $[-8,6]$.

In the *direct* approach, the empirical probability density function of the normalized data was estimated using a Gaussian kernel with reflective boundaries, and the corresponding probabilities were used to evaluate $H_\alpha^{(\text{dir})}$ while in the *symbolic* approach, the same data were transformed into symbolic sequences. Contiguous symbolic words of length $L_{\text{seq}} = 3$ were then used to construct the n-gram probability distribution.

Figure 4 compares the normalized Rényi entropy $H_\alpha$ obtained through the direct (left panels) and symbolic (right panels) approaches for both the pure and mixed surrogate series.

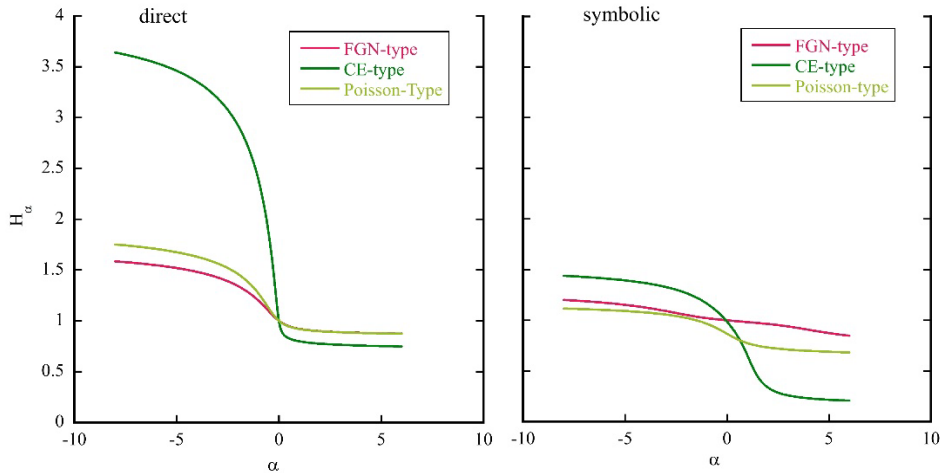

**Figure 4.** Normalized Rényi entropy $H_\alpha$ obtained from the direct (left) and symbolic (right) methods for the pure series: Poisson-type (light green), FGN-type (green), and CE-type (red).

In the pure cases (Poisson, FGN, and CE), the direct analysis highlights a strong contrast between Poisson and CE dynamics. The Poisson process yields an almost symmetric and weakly α-dependent entropy profile, consistent with an uncorrelated stochastic source. The FGN series, characterized by persistent long-range correlations ($H = 0.7$), exhibits a smoother but still monotonic decrease of $H_\alpha^{(\text{dir})}$ with α, reflecting moderate statistical broadening. In contrast, the CE series shows a pronounced asymmetry, with large $H_\alpha^{(\text{dir})}$ values for $\alpha < 0$ and a sharp decay for $\alpha > 0$, corresponding to the coexistence of rare, high-intensity events and recurrent low-activity states. The symbolic Rényi analysis confirms this differentiation: while Poisson and FGN signals produce nearly flat profiles, the CE process retains a marked asymmetry between the rare and frequent symbolic patterns, evidencing its intermittent grammar.

Figure 5 shows the Rényi entropies for the mixed processes, the direct Rényi curves converge toward a more compact shape, as the Poisson background partially suppresses the heavy tails of the CE distribution and reduces the amplitude heterogeneity induced by long-range correlations. Nonetheless, the symbolic representation reveals that temporal organization remains discriminative. The F-type series preserves a nearly uniform symbolic entropy, consistent with correlated but regular fluctuations, whereas C-type series displays a strong α-asymmetry, indicating that the sequence of events still bears the hallmark of crucial intermittency despite the added stochastic background. The fully mixed CF-type signal exhibits an intermediate behavior, suggesting that the fractional-Gaussian component regularizes, but does not eliminate, the intermittent symbolic structure. Altogether, these results demonstrate that direct Rényi entropy captures the statistical breadth of amplitude fluctuations, while the symbolic formulation isolates the intrinsic temporal constraints governing the emission dynamics.

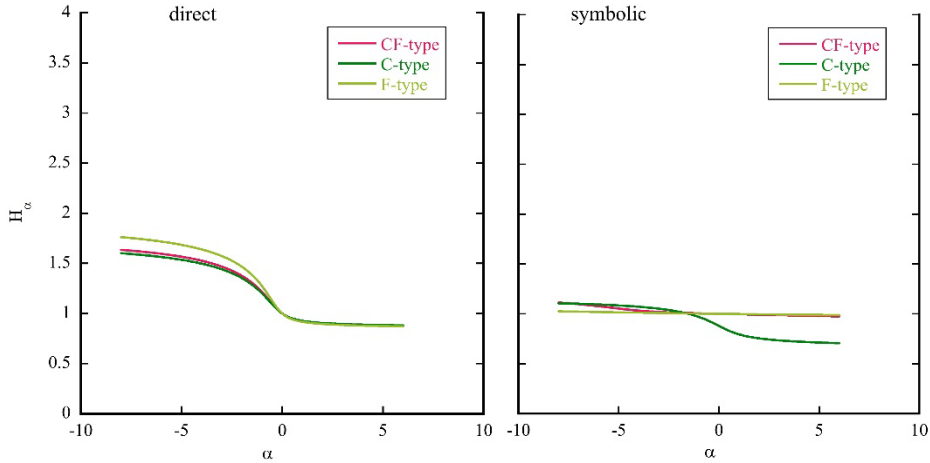

**Figure 5.** Normalized Rényi entropy $H_\alpha$ from the direct (left) and symbolic (right) methods for the mixed series: F-type (light green), C-type (green), and CF-type (red). The direct curves appear more compact due to the Poisson background, while in the symbolic analysis the C-type retains a pronounced α-asymmetry, and the CF-type shows an intermediate behavior between the intermittent C-type and the correlated F-type dynamics.

A notable difference emerges between the direct and symbolic Rényi analyses when comparing pure and mixed series. In the direct approach, the addition of a Poisson component to a crucial-event signal markedly reduces the high-α asymmetry observed in the pure CE case. This effect arises because the kernel-based estimation of the continuous probability density function tends to "fill in" the long tails introduced by intermittent bursts with the dense fluctuations of the Poisson background, thereby compressing the overall distribution and lowering $H_\alpha^{(dir)}$ for negative α. In contrast, the symbolic formulation remains sensitive to the temporal grammar of the process. The same Poisson admixture that regularizes the amplitude distribution does not erase the sequential imprint of intermittency, and the corresponding $H_\alpha^{(seq)}$ retains the characteristic high–low asymmetry between rare and frequent symbolic patterns. This divergence underscores the complementary nature of the two estimators: the direct Rényi entropy captures statistical broadening in the amplitude domain, while the symbolic one isolates the dynamic constraints emerging from temporal organization.

The results of the Rényi entropy analysis are fully consistent with the scaling behaviour revealed by the DEA. Both approaches point to a clear hierarchy among the underlying dynamical regimes: Poisson-like emission corresponds to a memoryless stochastic process ($\eta \approx 0.5$, nearly flat $H_\alpha$), while the F-type and C-type series display, respectively, the signatures of long-range correlation and crucial intermittency. The mixed CF-type signal retains intermediate features in both analyses, confirming that fractional-Gaussian correlations partially regularize the intermittency induced by crucial events without suppressing it entirely. In this sense, the Rényi framework complements the DEA by extending the characterization from temporal scaling to the full statistical structure of the fluctuations, bridging dynamical memory and amplitude heterogeneity within a unified entropic description.

To complete the analysis, we now turn to the Detrended Fluctuation Analysis and its multifractal extension, which are designed to extract correlation exponents and multifractal signatures even in the presence of nonstationarity. DFA quantifies the global persistence of fluctuations through the Hurst exponent $H$, while MFDFA extends this framework by resolving how fluctuations of different amplitude scales across time, yielding the generalized Hurst spectrum $h(q)$ and the associated multifractal singularity distribution $f(\alpha)$. In this way, DFA/MFDFA provide the temporal-domain counterpart to the entropy-based characterization obtained above.

Figure 6 shows the MFDFA spectrum of the different surrogate time series.

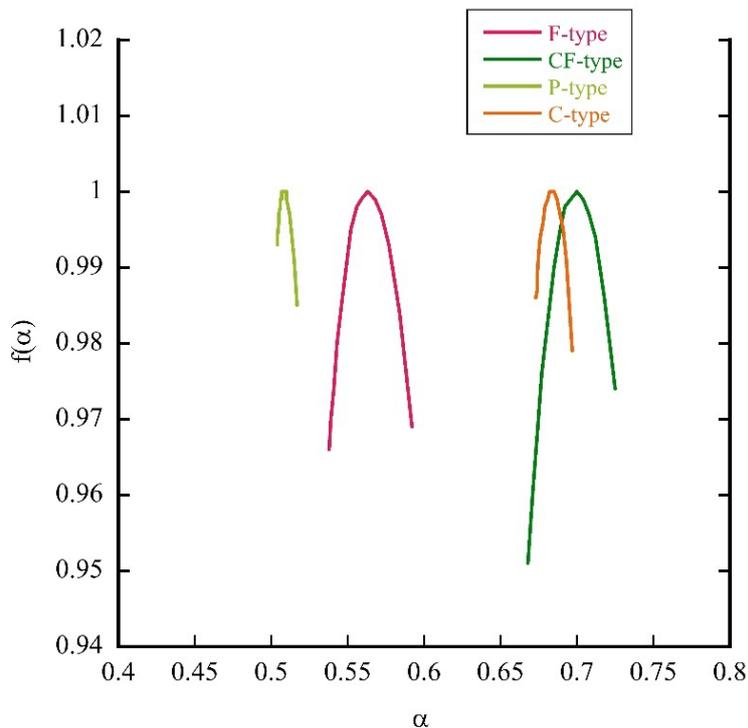

**Figure 6.** Shown are the spectra for Poisson noise (P-type, light green), fractional Gaussian noise (F-type, red), crucial-event dynamics (C-type, orange), and the mixed process (CF-type, dark green). The Poisson series displays a narrow monofractal peak near $\alpha \approx 0.50$, while F-type data broaden around $\alpha \approx 0.57$, consistent with persistent Gaussian correlations. C-type and CF-type series exhibit wider spectra shifted toward higher $\alpha$, reflecting intermittent, renewal-driven multifractality. Together, the spectra summarize the progression from purely random to increasingly organized and bursty dynamics.

The multifractal spectra obtained via MFDFA show a clear differentiation among the surrogate time series, consistent with the trends revealed by DEA and Rényi entropy. The Poisson process exhibits a narrow spectrum centred around $\alpha \approx 0.5$, consistent with monofractal, memoryless dynamics. The addition of fractional Gaussian noise (F-type) shifts the peak to $\alpha \approx 0.58$ and increases the spectral width, reflecting long-range correlations and a weak multifractal signature associated with fractional Brownian fluctuations.

Crucial events (C-type) lead to a pronounced rightward shift of the spectrum (peak around $\alpha \approx 0.68$), together with increased width, indicating intermittent dynamics with heterogeneous scaling exponents. When both mechanisms coexist (CF-type), the spectrum becomes broader and slightly further shifted ($\alpha \approx 0.70$), showing that fractional correlations reinforce, rather than mask, the intermittent structure introduced by crucial events.

Overall, MFDFA confirms the hierarchy of dynamical regimes detected by DEA and Rényi analysis:

$$\text{Poisson} < \text{Poisson+FGN} < \text{Poisson+CE} < \text{Poisson+CE+FGN}$$

in terms of complexity and multifractal richness. This convergence across independent statistical indicators supports the robustness of our interpretation.

To quantify the degree of multifractality in the analysed time series we employed two complementary indicators derived from the Multifractal Detrended Fluctuation Analysis framework. First, we computed the spectral width

$$\Delta\alpha = \alpha_{\max} - \alpha_{\min},$$

which measures the range of singularity strengths in the multifractal spectrum $f(\alpha)$. For purely monofractal processes (e.g., Poisson or fractional Gaussian noise), $\Delta\alpha$ approaches zero, while broader spectra indicate increasing heterogeneity of scaling exponents and thus stronger multifractality [41,46].

In addition, we used the asymmetry index

$$\Delta h = h(q_{\min}) - h(q_{\max}),$$

where $h(q)$ is the generalized Hurst exponent and $q_{\min}$ and $q_{\max}$ are the extreme $q$ values used in MFDFA. For monofractal signals, $h(q)$ is constant and therefore $\Delta h \approx 0$.

Table 4 reports the average Hurst exponent obtained from DFA, together with the multifractal spectral width and asymmetry index coming from the MFDFA analysis, for the four classes of surrogate time series. The Hurst exponent reported in Table 4 was obtained from the DFA analysis as the average slope of the log–log plot of the fluctuation function $F(s)$ versus the scale $s$. Only the stable scaling region was included in the fit, while the largest-scale points—where the number of segments becomes insufficient and fluctuations dominate—were excluded to avoid statistical bias.

|        | $H$  | $\Delta\alpha$ | $\Delta h$ |
|--------|------|----------------|------------|
| P-type | 0.50 | 0.013          | 0.006      |
| C-type | 0.69 | 0.024          | 0.013      |
| F-type | 0.54 | 0.072          | 0.036      |
| CF-type| 0.69 | 0.070          | 0.034      |

**Table 4.** Hurst exponent $H$ from DFA, multifractal spectrum width $\Delta\alpha$, and asymmetry $\Delta h$ obtained from MFDFA for the four surrogate series. Poisson dynamics yield the expected monofractal behavior ($H \approx 0.50$) with minimal spectral width, while crucial events induce strong persistence and modest multifractality. The FGN-modulated process exhibits mild persistence and enhanced intermittency due to its interaction with discrete noise, and the combined CE+FGN signal shows the strongest joint signature of persistence and multifractality.

The DFA/MFDFA indicators align with the expected properties of the surrogate dynamics. The Poisson process yields $H \approx 0.50$ and an extremely narrow multifractal spectrum, consistent with memoryless monofractal fluctuations. The C-type and CF-type series display $H \approx 0.69$, reflecting the persistent long-range dependence imposed by crucial events. The F-type signal, a Poisson process modulated by fractional Gaussian noise, exhibits a milder persistence ($H \approx 0.54$), lower than the nominal $H = 0.70$ of the generating fractional Gaussian noise. This reduction is expected, as the discrete Poisson component partially masks long-range correlations. Importantly, pure FGN is essentially monofractal ($\Delta\alpha \approx 0.009$), as predicted by Gaussian self-similarity theory: Gaussian processes with stationary increments scale uniformly and do not generate multifractality in the absence of intermittency [47]. However, when FGN is superimposed on discrete Poisson events, the resulting departure from Gaussian behavior produces heterogeneous local fluctuations, especially in the distribution tails. This interaction between long-range correlated fluctuations and impulsive shot noise broadens the multifractal spectrum by introducing local intermittency that is absent in pure FGN [47,48].

The pure crucial-events series exhibits strong persistence ($H \approx 0.69$) but only moderate multifractality, confirming that heavy-tailed waiting-time statistics enhance temporal memory without necessarily diversifying local scaling exponents. In contrast, the mixed CF-type dynamics, which combine scale-free waiting times and correlated amplitude modulation, yield both strong persistence and the widest multifractal spectrum, demonstrating that multifractality arises most prominently when power-law temporal organization and correlated fluctuation amplitudes act jointly.

The tail statistics and Fano factors support the same hierarchy revealed by DFA/MFDFA. Pure Poisson dynamics are featureless, monofractal, and memoryless, while crucial events introduce long-range temporal dependence and heavy-tailed fluctuations but only moderate multifractal spreading. Fractional Gaussian noise alone remains nearly monofractal, and a broad multifractal spectrum emerges only when long-range Gaussian correlations interact with discrete burst-like dynamics. Accordingly, the mixed CE+FGN series exhibits the strongest intermittency and the widest range of scaling exponents. Overall, the DFA/MFDFA results consistently mirror the trends revealed by DEA and Rényi analysis, confirming that crucial-event and hybrid dynamics give rise to persistent and multifractal behavior, whereas Poisson statistics provide a memoryless monofractal baseline.

Across the four surrogate models, the combined results of DEA, Rényi entropy, DFA/MFDFA, and tail-statistics analysis converge to a coherent and internally consistent picture of their dynamical structure. As expected, the Poisson benchmark exhibits purely stochastic behavior: DEA yields $\eta \approx$

0.5, Rényi spectra remain essentially flat and order-independent, DFA gives $H \approx 0.50$, and the multifractal width is negligible. Likewise, tail indicators confirm the absence of long-range memory, heavy tails, or hierarchical scaling, defining a clean monofractal, memoryless baseline.

Introducing crucial events fundamentally alters the statistical picture. DEA departs from the Brownian value, Rényi entropy develops a marked asymmetry in α, DFA yields $H \approx 0.69$, and tail metrics reveal heavy–tailed waiting times and super-Poissonian fluctuations. These features reflect renewal-type persistence and aging induced by power-law inter-event statistics, rather than continuous long-range correlations. Consistently, the multifractal width remains moderate: heavy-tailed waiting times alone generate strong intermittency and deviation from Poisson statistics, but do not create a broad hierarchy of scaling exponents. Fractional Gaussian noise exhibits complementary behavior. Although generated with H=0.7, the pure FGN series is nearly monofractal, as expected for Gaussian self-similar dynamics. DFA correctly recovers the imposed persistence, but Rényi and MFDFA show only weak multifractality. When FGN is combined with discrete shot-like fluctuations (F-type surrogate), mild multifractality emerges DEA shifts only slightly above the Poisson value, Rényi spectra acquire curvature, and MFDFA detects a non-zero but limited spectral width. Tail statistics confirm moderate intermittency and weaker heavy-tail behavior compared to crucial-event dynamics.

The richest dynamical pattern appears in the mixed CF-type series. Here, DEA, Rényi entropy, and MFDFA consistently reveal renewal-driven persistence together with the broadest multifractal spectrum, while tail analysis shows the largest intermittency index. This demonstrates that the most complex regime arises not from a single mechanism, but from the synergistic interplay between heavy-tailed renewal structure and genuine long-range stochastic fluctuations.

This surrogate-based benchmark establishes a clear hierarchy, Poisson baseline, memory-driven CE dynamics, Gaussian-correlated modulation, and a maximally complex mixed regime, providing a rigorous interpretive frame for the analysis of real photon emission data.

With the surrogate benchmark established, we next analyze real counting sequences acquired under controlled experimental conditions. These include the intrinsic dark-count signal of the detector and photon counts generated by a weak coherent laser field. This provides a physically grounded reference against which future biological signals can be interpreted.

### 4.0. Real Data Analysis

The analysis then extends to real photocounting data, beginning with measurements of the detector dark signal acquired at 21°C, followed by photon-count statistics obtained from an attenuated laser source used as a coherent-light reference. The experimental setup and acquisition procedure follow those described in our previous works [17,25], ensuring continuity and comparability with established detection protocols. The laser emission at 658 nm (few mW output power) was attenuated by an optical density (OD) filter with OD = 3.3, corresponding to an attenuation of approximately $2 \times 10^3$. After attenuation, the residual optical power coupled into the measurement chamber was in the sub-µW range. The attenuated beam was delivered through an optical fiber inserted through a sealed aperture in the measurement chamber floor, in the position typically occupied by the Petri dish.

Figure 7 shows the experimental data of dark count and laser emission.

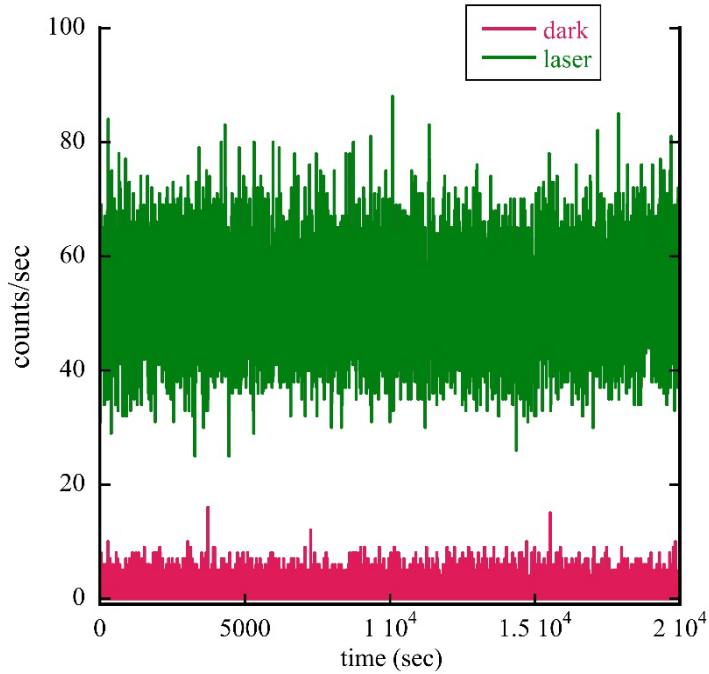

**Figure 7.** Photomultiplier count time series acquired in darkness (lower, red) and under attenuated laser illumination (upper, green). The plot shows the first 20 000 s of acquisition; the complete analyzed series spans 100 000 s at 1 Hz sampling.

The lower trace (red curve) corresponds to the photomultiplier dark counts, while the upper trace (green curve) represents the photon count obtained from the appropriately attenuated laser emission. Table 5 summarizes the main statistical properties of these time series together with the coefficients extracted from the tail–distribution analysis.

|  | $<m>$ | $\sigma$ | Skewness | $M_{exp}(q)$ | $ES_{exp}(q)$ | Fano factor |
|---|---|---|---|---|---|---|
| dark | 1.50 | 1.50 | 1.467 | 0.100 | 4.78 | 1.503 |
| laser | 52.9 | 7.79 | 0.155 | 0.111 | 65.52 | 1.148 |

**Table 5.** Statistical indicators for the dark–noise and attenuated–laser photon-count time series. The dark counts exhibit super-Poissonian fluctuations and strong asymmetry, while the attenuated laser is near-Poissonian with weak deviation from shot-noise statistics.

The statistical properties of the dark–noise and attenuated laser photon count sequences clearly distinguish their physical origins. The dark signal exhibits $\langle m \rangle \approx 1.5$ counts/s with a standard deviation comparable to the mean, corresponding to a variance significantly larger than the mean and a Fano factor $F \approx 1.50$. This super-Poissonian behaviour reflects the presence of sporadic burst-like activation events in the detector, consistent with electronic noise spikes. The strong positive skewness ($\approx 1.47$) and moderate heavy-tail exponent ($M_{exp}(q) = 0.10$) further confirm intermittent fluctuations and non-Gaussian statistics in the dark counts.

Conversely, the attenuated laser series displays $\langle m \rangle \approx 52.9$ counts/s and $\sigma \approx 7.8$, yielding a Fano factor only slightly above unity ($F \approx 1.15$), as expected for a nearly coherent light source under strong attenuation. The distribution is nearly symmetric and only weakly heavy-tailed ($M_{exp}(q) = 0.11$), consistent with quasi-Poissonian photon statistics. The small excess variance likely arises from residual instrumental noise and coupling fluctuations in the fiber-coupled delivery system.

In full analogy with the analysis performed on the surrogate series, we now apply DEA, with and without stripes, to the experimental dark-noise and attenuated-laser time series. This allows us to directly compare the scaling properties of real photon-counting data with the benchmark behaviors established in the controlled synthetic cases. In table 6 the scaling factor $\eta$ is reported for DEA with and without stripes.

|  | $\eta$ (no-str) | $\eta$ [$\Delta_s = 3 \cdot \sigma$] |
|---|---|---|
| dark | 0.51±0.01 | 0.52±0.01 |
| laser | 0.50±0.02 | 0.51±0.02 |

**Table 6.** Scaling exponents $\eta$ obtained from DEA analysis of dark counts and attenuated laser emission, with and without stripes ($\Delta_s = 3 \cdot \sigma$). Both datasets exhibit $\eta \approx 0.5$, confirming Poisson-like statistics and absence of temporal correlations.

The DEA results for both the dark counts and the attenuated laser emission confirm the expected behavior of purely stochastic photon processes. In the absence of structured dynamics, both signals yield scaling exponents consistent with the theoretical Poisson value, $\eta \approx 0.5$. For the dark signal, $\eta = 0.51 \pm 0.01$ indicates that the apparent super-Poissonian dispersion observed in the raw statistics does not correspond to temporal correlations but rather reflects the intrinsic excess variance of the detector noise. Similarly, the attenuated laser, expected to generate ideal shot noise, exhibits $\eta = 0.50 \pm 0.02$, validating the photon-counting regime of the experimental setup.

Importantly, the agreement between the standard DEA and the striped formulation demonstrates that the stripes procedure does not introduce artifacts when the signal lacks burst-like structure or long-range dependence. The nearly identical scaling values, $\eta(no-str) \approx \eta(stripes)$, highlight that no hidden intermittency or memory is present in either dataset, and confirm the method's robustness in the canonical Poisson regime.

As expected, the attenuated laser signal exhibits Poisson statistics, with $\eta \approx 0.5$ in both DEA configurations. This behavior reflects the quantum shot-noise regime of coherent light: photon arrivals are independent events with a constant emission probability per unit time, leading to equality between variance and mean and the absence of temporal correlations [26,49]. In photon-counting experiments, the second-order correlation at zero delay, $g^{(2)}(0)$, characterizes the degree of photon arrival correlations. For stationary counting processes, a useful relation links $g^{(2)}(0)$ to the Fano factor $F = \sigma^2/\langle m \rangle$

$$g^{(2)}(0) \approx 1 + \frac{F-1}{\langle m \rangle}$$

For a Poisson process, $F = 1$ and therefore $g^{(2)}(0) = 1$. Values $g^{(2)}(0) > 1$ indicate super-Poissonian statistics (photon bunching or excess noise), while $g^{(2)}(0) < 1$ reveals sub-Poissonian emission (photon antibunching). Although $g^{(2)}(0)$ is not directly measured here, the statistical results obtained for the attenuated laser signal, namely $\eta \approx 0.5$, near-unity Fano factor, and absence of heavy-tailed fluctuations, are fully consistent with the shot-noise statistics of coherent light, for which $g^{(2)}(0) = 1$. In this sense, the laser dataset behaves as an experimental realization of an ideal Poisson process.

To further assess the statistical structure of the experimental time series we now apply the Rényi entropy analysis, using both direct and symbolic estimators as in the surrogate-series study. This allows us to test whether the dark-noise and laser signals exhibit any deviation from the reference Poisson behavior in terms of amplitude statistics or temporal patterning. Fig. 8 shows the Rényi entropy curves for the dark counts and attenuated laser signals. The Rényi entropy of an ideal Poisson process is also included for comparison.

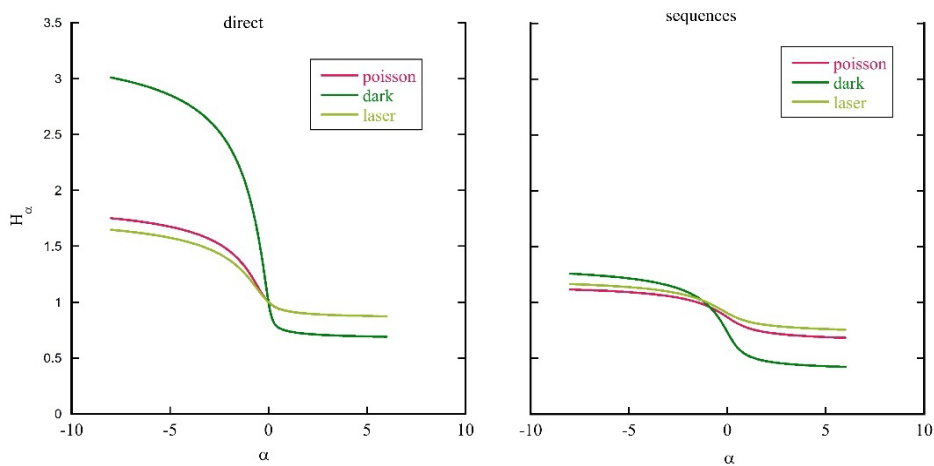

**Figure 8.** Rényi entropy for dark counts and attenuated laser signals, compared with an ideal Poisson reference. Both datasets follow the Poisson trend, with a slight deviation for dark noise at negative α due to super-Poissonian fluctuations. For the laser data, the symbolic alphabet size and sequence length were $K_{\text{seq}} = 65$ and $L_{\text{seq}} = 2$, respectively, while for the dark data they were $K_{\text{seq}} = 22$ and $L_{\text{seq}} = 3$.

Both the dark–count data and the attenuated-laser series closely follow the Rényi-entropy trend of an ideal Poisson process, confirming the absence of structured temporal organization and the essentially memoryless nature of photon arrival statistics in both cases. The attenuated laser overlaps the Poisson reference across the full α-range for both direct and symbolic estimators, as expected for coherent light in the shot-noise regime. The dark signal remains similarly consistent with Poisson behavior in the symbolic representation; however, the direct Rényi estimator exhibits a noticeable deviation for negative α, where the entropy becomes sensitive to rare events and low-probability amplitude bins. This effect does not indicate correlations but instead reflects the highly discrete and low-mean character of photomultiplier dark noise, with a large fraction of zero-count intervals and occasional electronic spikes that disproportionately influence the $\alpha < 0$ sector. The symbolic estimator effectively regularizes this distributional roughness and recovers the expected near-Poisson trend, corroborating that both experimental series behave as uncorrelated counting processes.

This behavior is fully consistent with the DEA results, where both dark and laser signals exhibited the Poisson scaling exponent $\eta \approx 0.5$ with and without stripes. Taken together, the Rényi and DEA analyses confirm that neither dataset displays temporal memory nor anomalous scaling, reinforcing their role as reliable Poissonian reference signals.

To complete the comparison between experimental photon–count data and the ideal Poisson benchmark, we now examine the same dark-noise and attenuated-laser series using DFA and MFDFA.

While DEA and Rényi entropy respectively probe temporal scaling and the structure of amplitude fluctuations, DFA provides a direct estimate of long-range correlation through the Hurst exponent, and MFDFA extends this evaluation to possible multifractal organization. Applying these methods to experimental datasets allows us to verify whether any residual dependence or intermittency exists beyond the purely Poissonian behavior identified by the previous analyses. The results of MFDFA analysis are reported in Fig.9

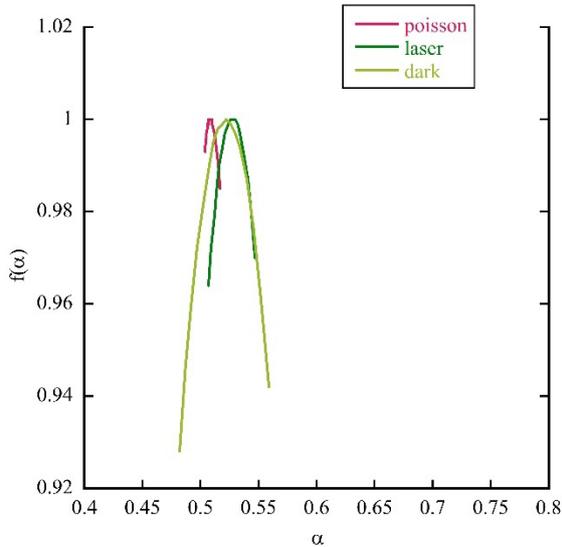

**Figure 9.** Multifractal spectra $f(\alpha)$ for the experimental dark-count series (light green), attenuated-laser emission (green), and a reference Poisson process (red). All three spectra collapse onto a narrow curve centered at $\alpha \approx 0.50$, with comparable spectral width and shape, indicating the absence of multifractality and confirming Poisson-like temporal statistics for both experimental signals.

The multifractal spectra obtained from the experimental dark-count and attenuated-laser series are nearly indistinguishable from that of a synthetic Poisson process. All three traces show a narrow $f(\alpha)$ curve centered at $\alpha \approx 0.50$, with a spectral width $\Delta\alpha$ of only a few percent. This indicates the absence of significant multifractality and confirms that both dark noise and the attenuated laser emission behave as nearly ideal Poisson point processes at all resolved scales. No broadening of the spectrum or asymmetry, typical signatures of temporal intermittency or amplitude heterogeneity, is observed, in agreement with the Renyi and DEA results and with the shot-noise nature of coherent laser light and PMT dark emission. The DFA/MFDFA indicators are consistent with Poisson-like dynamics. For the experimental series we obtain $\Delta\alpha = 0.077$ and $\Delta h = 0.035$ (dark values), and $\Delta\alpha = 0.041$ and $\Delta h = 0.019$ (laser values), with mean DFA Hurst exponents close to $H \approx 0.5$ in both cases. As expected, a synthetic Poisson reference yields smaller multifractal widths than these values; the slight broadening observed in the experimental data is compatible with instrumental noise and residual count-rate fluctuations rather than genuine multifractality.

## 5.0. Conclusions

The present work has introduced and systematically compared four complementary analytical frameworks, Distribution Entropy Analysis (DEA), Rényi entropy, Detrended Fluctuation Analysis (DFA), and its multifractal extension (MFDFA), to assess their ability to detect and characterize non-Poissonian dynamics and long-range organization in photon counting time series. By benchmarking the methods on well-controlled surrogate signals and subsequently applying them to experimental photon-count data, we have established a coherent methodology for distinguishing trivial stochastic processes from structured, self-organized emission dynamics such as those hypothesized in biological ultra-weak photon emission.

Each method contributed unique information. DEA linked temporal scaling to the information-theoretic entropy growth, placing emphasis on temporal diffusion. Rényi entropy dissected probability structure across the full spectrum of fluctuations, with symbolic encoding providing sensitivity to ordering patterns invisible to amplitude-based estimators. DFA isolated long-range correlations under nonstationarity, while MFDFA quantified the diversity of scaling exponents. In addition, the analysis of distribution tails and Fano statistics provided an independent check on fluctuation asymmetry and dispersion properties, reinforcing the dynamical inferences drawn from entropy- and scaling-based methods. Despite their different conceptual formulations, all indicators—including tail exponents—converged toward the same classification of surrogate dynamics, demonstrating robustness and methodological complementarity.

When applied to real photon-count data from the detection apparatus—dark noise and attenuated coherent laser emission—the four techniques consistently identified Poisson-like behavior. Both sources yielded $H \approx 0.5$, narrow multifractal spectra, and Rényi profiles close to the theoretical Poisson reference, together with tail exponents and Fano factors consistent with uncorrelated emission statistics. Minor deviations, particularly in the dark counts, likely stem from instrumental fluctuations rather than intrinsic dynamics, confirming the reliability and sensitivity of the approach. The absence of anomalous scaling signatures in these controlled tests establishes a crucial baseline for future biophoton studies, ensuring that any departure from Poisson statistics observed in living systems cannot be attributed to detector noise or trivial coherent input.

Taken together, these results demonstrate that the combined use of DEA, Rényi entropy, DFA, MFDFA, and tail analysis constitutes a powerful and internally coherent toolkit for detecting subtle signatures of self-organization, intermittency, and memory in photon emission. The approach is particularly suited to the study of biological ultra-weak photon signals, where stochastic and dynamical components coexist and where departures from equilibrium-like emission may reflect metabolic coherence, adaptive control, or critical-like dynamics. The methodology established here thus provides both a conceptual foundation and a practical reference framework for the quantitative exploration of photonic coherence and complexity in living systems. Future work will extend this pipeline to biological samples, with the goal of mapping how dynamical signatures evolve during physiological transitions, stress responses, and macroscopic growth processes. Beyond mere statistical classification, this framework opens the possibility of linking biophoton emissions to functional biological states, enabling rigorous tests of hypotheses concerning long-range cellular communication, coherent biochemical organization, and quantum-like photonic processes in living matter. By integrating statistical physics, information theory, and photon counting, this research program aims to clarify whether biophoton emission encodes functional organizational principles and whether such photonic signatures may serve as markers of emergent biological coherence.

**Acknowledgment:** The authors gratefully acknowledge I. Davoli, R. Francini, F. De Matteis, and A. Clozza for their indispensable contribution to the acquisition of the experimental data. We further thank E. Pace and C. Curceanu for insightful discussions and for their encouragement throughout the development and writing of this work.